\documentclass[review]{elsarticle}
\usepackage{lineno}
\usepackage{amssymb}
\usepackage[figuresright]{rotating}

\begin{document}
\begin{frontmatter}
\title{Interplay of fission modes in mass distribution of light actinide nuclei $^{225, 227}$Pa}

\author[iuac]{ R. Dubey}
\author[iuac]{ P. Sugathan\corref{cor}}
\cortext[cor]{Corresponding author Tel. : +91 11 26893955, Fax : +91 11 26893666}
\ead{sugathan@iuac.res.in}
\author[iuac]{ A. Jhingan}
\author[PU]{ Gurpreet Kaur}
\author[iuac]{ Ish Mukul\fnref{now1}}
\fntext[now1]{Present address: Department of Particle Physics and Astrophysics, Weizmann Institute of Science, Rehovot 76100, Israel}
\author[iuac]{G. Mohanto\fnref{now2}}
\fntext[now2]{Present address: Nuclear Physics Division, Bhabha Atomic Research Centre, Mumbai 400085, India }
\author[iuac]{ D. Siwal\fnref{now3}}
\fntext[now3]{Present address: Department of Chemistry and Center for Exploration of Energy and Matter, Indiana University, 2401 Milo B. Sampson Lane, Bloomington, IN 47408, USA}
\author[iuac]{ N. Saneesh}
\author[iuac]{ T. Banerjee}
\author[PU]{ Meenu Thakur}
\author [PU]{ Ruchi Mahajan}
\author[DU]{ N. Kumar}
\author[SINP]{ M. B. Chatterjee} 
\address[iuac]{Inter University Accelerator Centre, Aruna Asaf Ali Marg, New Delhi 110067, India}
\address[PU]{Department of Physics, Panjab University, Chandigarh 160014, India}
\address[DU]{Department of Physics and Astrophysics, University of Delhi, New Delhi 110007, India}
\address[SINP]{Saha Institute of Nuclear Physics, Kolkata, West Bengal 700064, India}

\begin{abstract}
Fission-fragment mass distributions were measured for $^{225, 227}$Pa nuclei formed in fusion reactions of $^{19}$F + $^{206, 208}$Pb around fusion barrier energies. Mass-angle correlations do not indicate any quasi-fission like events in this bombarding energy range.   Mass distributions were fitted by Gaussian distribution and mass variance extracted. At below-barrier energies, the mass variance was found to increase with decrease in energy for both nuclei. Results from present work were compared with existing data for induced fission of $^{224, 226}$Th and $^{228}$U around barrier energies. Enhancement in mass variance of $^{225, 227}$Pa nuclei at below-barrier energies shows evidence for presence of asymmetric fission events mixed with symmetric fission events. This is in agreement with the results of mass distributions of nearby nuclei $^{224, 226}$Th and $^{228}$U where two-mode fission process was observed. Two-\underline{}mode feature of fission arises due to the shell effects changing the landscape of the potential-\emph{}energy surfaces at low excitation energies. The excitation-energy dependence of the mass variance gives strong evidence for survival of microscopic shell effects in fission of light actinide nuclei $^{225, 227}$Pa with initial excitation energy $\sim 30 - 50$ MeV.
\end{abstract}
\end{frontmatter}

\section{Introduction}
 The mechanism of mass division in fission of atomic nuclei has been an intriguing problem in nuclear physics for several years. Generally, it is observed that asymmetric mass division is predominant in spontaneous fission or low energy induced fission of actinide nuclei \cite {Turkevich} whereas nuclei  around $^{208}$Pb fission mainly to symmetric mass division \cite{Fairhall}. Some nuclei in the region of $^{228}$Ra shows three-humped structure of the mass distribution  reflecting contributions from both symmetric and asymmetric mass components \cite{Jensen, Konecny}. With growing excitation energy of the fissioning nucleus, the asymmetric mass distribution changes to  symmetric Gaussian distribution. These characteristics are explained according to the concept of independent fission channels (modes)  which corresponds to specific valleys in the potential-energy surface (PES) of the fissioning nucleus \cite{Turkevich,Paskevich71, Brosa}. The origin of these valleys has been attributed to shell structures modifying the potential energy landscape of the deformed system \cite{Moller74,Wilkins,Mustafa, V. Pokrovsky}. More recent  calculation of five-dimensional PES by M\"{o}ller  \textit{et al.} well predicts the presence of two deformation paths; one path  with elongated scission configuration leading to symmetric mass division and the other shell influenced more compact scission configuration leading to asymmetric mass division \cite {Moller01}.  Two-mode nature of fission phenomenon has been observed in induced fission of light and heavy actinides \cite {Balagna,Hoffman, Hulet, Ohtsuki, Nagame}. The effects of A, Z and characteristics of final mass division from each fission mode has been studied extensively. The seminal work by Schmidt \textit{et al.} on the charge distribution of fission fragments(FF) in Coulomb fission of 70 neutron deficient isotopes (Z = 85-92) suggested that with increasing nucleon number, a transition takes place from the symmetric to the  asymmetric fission mode around  mass A$\approx$ 226 in this region of nuclei \cite{Schmidt}.
     
 The variance of the mass distribution ($\sigma_M^2$) and its dependence on excitation energy have been used as sensitive probe to study the dynamics of fission in heavy ion (HI) induced reactions \cite{Ogn85}. Typically mass yields from complete fusion fission are found to be  symmetric with its variance  changing linearly with temperature.  Presence of non-compound process such as quasi-fission(QF) broadens the mass distribution.  For example, in reaction induced by $^{16}$O projectile on deformed target $^{238}$U at energies below fusion barrier, sudden increase in  $\sigma_M^2$ have been observed and attributed  to the effects of QF \cite {Banerjee}. On the contrary, FF mass distribution in the reaction $^{12}$C + $^{235}$U showed substantial mass-asymmetric component with increasing yields at low excitation energies which was attributed to shell effects \cite {Khuyagbaatar}. Though theoretically  QF has been predicted in reactions with  Z$_{p}$Z$_{T}$ (projectile target charge product) $\ge$1600, recent measurement showed evidence of QF even in less fissile systems with Z$_{p}$Z$_{T} <$ 800 \cite { Banerjee,T.K.Ghosh, Hinde}. It is possible that the increased mass variance $\sigma_M^2$  at lower energies observed in some actinide nuclei could also be due to the asymmetric fission components manifesting at low excitation energies.  Hence, it is important to distinguish the contribution of QF and asymmetric fission events in the mass distribution of fragments in heavy ion induced reactions. 
  
  In this letter, we report on experimental measurements of mass distribution of FF from reactions $^{19}$F + $^{206,208}$Pb forming $^{225, 227}$Pa compound nuclei (CN) over a range of excitation energies ($E_{CN}^*$ = 30-50 MeV). The selected reaction has low Z$_{p}$Z$_{T}$ (736) and the fission fragment angular distribution data already exists for $^{19}$F + $^{208}$Pb system \cite {Hinde1}. By choosing the targets to be spherical, the influence of deformation effects can be eliminated. It is worth to mention that for the present system $^{19}$F + $^{206, 208}$Pb under study, entrance channel mass-asymmetry and Coulomb interaction  would be almost identical for both these systems. This enabled us to study  the dynamics of the fission of the two $^{225,227}$Pa nuclei almost under identical conditions matching their angular momentum and excitation energies. It is expected that the mass distribution of these systems follows the predictions of standard normal fusion-fission dynamics within the range of measured energies. As these two nuclei differ only by two neutrons and fall on either side of the transition mass limit A$_{CN}$ =226, the two reactions can determine if the neutron number make any observable differences on the width of their mass distributions that might be expected due to different fission modes existing in this mass region.
   
\section{Experimental setup}
The experiment was carried out using $^{19}$F beams from the 15UD Pelletron accelerator at Inter University Accelerator Centre, New Delhi, India. Pulsed beam with width of $\sim 1.5$ $ns$ and separation of 250 $ns$ was used to bombard isotopically enriched $^{206, 208}$Pb targets of $\sim$ 110 $\mu$g/cm${^2}$ thickness deposited  on $\sim20$ $\mu$g/cm${^2}$ thick $^{12}$C backings. The experiment was performed at different beam energies varying from 87 MeV to 120 MeV choosing the energy steps to form compound nuclei with similar excitation energies. The complimentary fission fragments were detected in coincidence using two large-area position-sensitive multi-wire proportional counters (MWPCs) of dimension $20cm \times 10 cm$ positioned in the forward and backward hemispheres \cite {Jhingan}. They were mounted on two arms of the 1.5 m diameter general-purpose scattering chamber. The detectors were operated with isobutane gas at two Torr pressure. Two Si surface-barrier detectors mounted at angle of $\pm$10${^\circ}$ with respect to the beam axis were used to monitor the beam energy and position of the beam at the center of the target during the experiment. Clean identification of fission events was made through the requirement of a kinematic coincidence between the fragments in the two detectors.

\section{Experimental results}
\begin{figure}[htb!]
\centering
\includegraphics[width=1\linewidth]{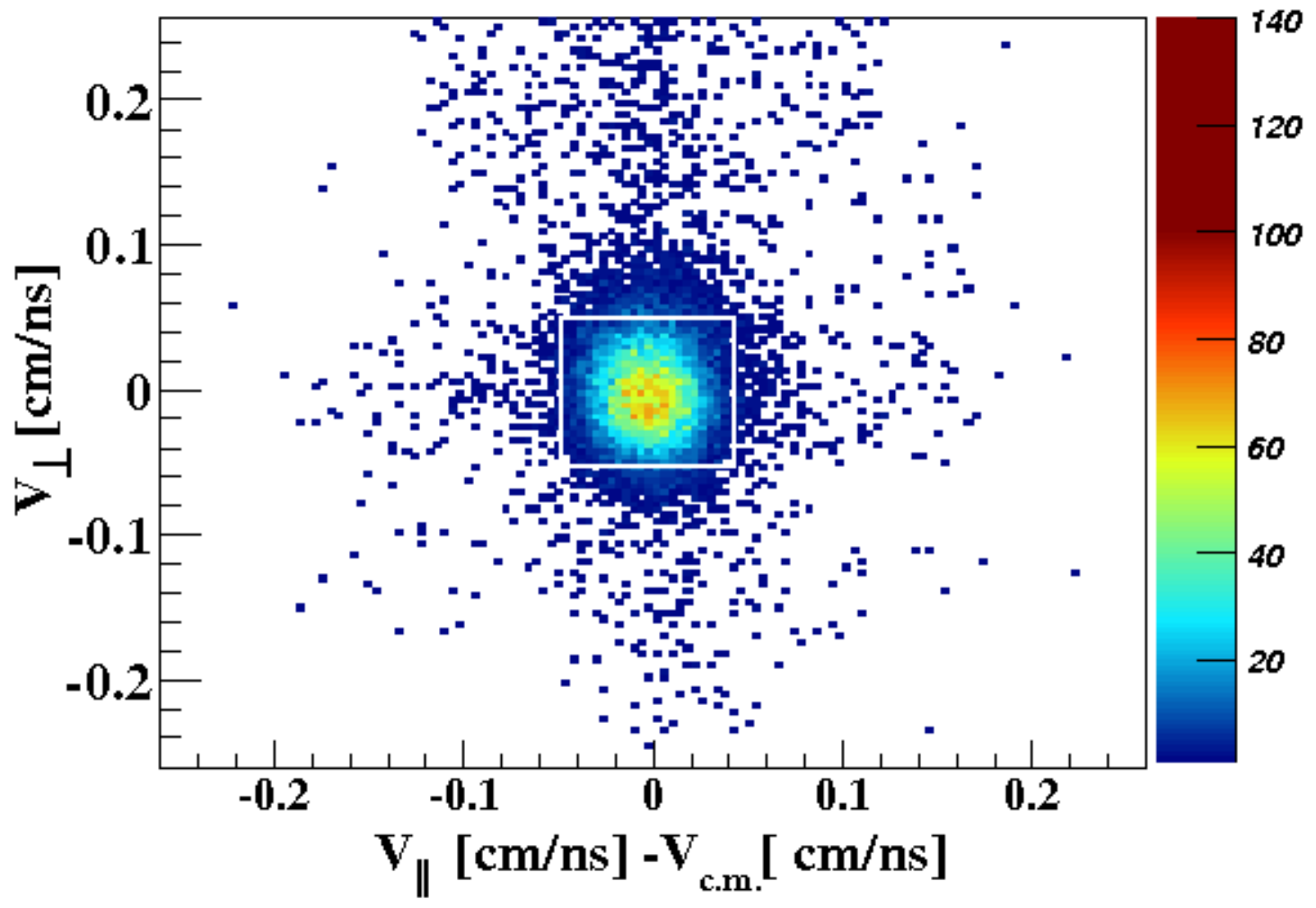}
\caption{\label{fig:Figure1} Measured distribution of velocity components of FF at 5\% below fusion barrier for the reaction $^{19}$F + $^{208}$Pb (beam energy 89 MeV). Full momentum transfer fission events are shown inside the rectangular box.}
\end{figure}

The fission fragments were separated from the elastic and quasi-elastic particles by time-of-flight and energy-loss signals in the MWPC. The data analysis has been performed following the velocity-reconstruction method given by \cite {Hinde96}. From the position and time distribution of fission fragments in MWPC, the polar angles (${\theta}$, ${\phi}$) and the velocity-vector components (parallel ($V_\parallel$) and  perpendicular ($V_\perp$) to beam direction)  of the fissioning nucleus were determined for each event. Fission events originating from complete fusion was selected by imposing the condition of full momentum transfer (FMT) of fission-like events using the correlation of velocity components. Fig. \ref{fig:Figure1} displays the spectra showing the correlation between  measured $V_\perp$ and $V_\parallel$ -$V_{c.m.}$ (where $V_{c.m.}$ is the center-of-mass velocity) for fission events from the reaction $^{19}$F + $^{208}$Pb at beam energy 89 MeV. The intense region around the velocity coordinates ($V_\parallel$ -$V_{c.m.}$,$V_\perp$)=(0,0) corresponds to the events originated from FMT fission. A software gate around these events, shown as white rectangle in the plot, was used in the analysis of mass-angle correlation and mass-ratio distribution. The mass ratio $M_R=\frac{m_1}{m_1+m_2}$ ($m_1$ and $m_2$ are two fragment masses) determined from the ratio of the velocities in the center-of-mass frame was used to generate mass distribution.

\begin{figure}[h!]	
\centering
\includegraphics[width=1\linewidth]{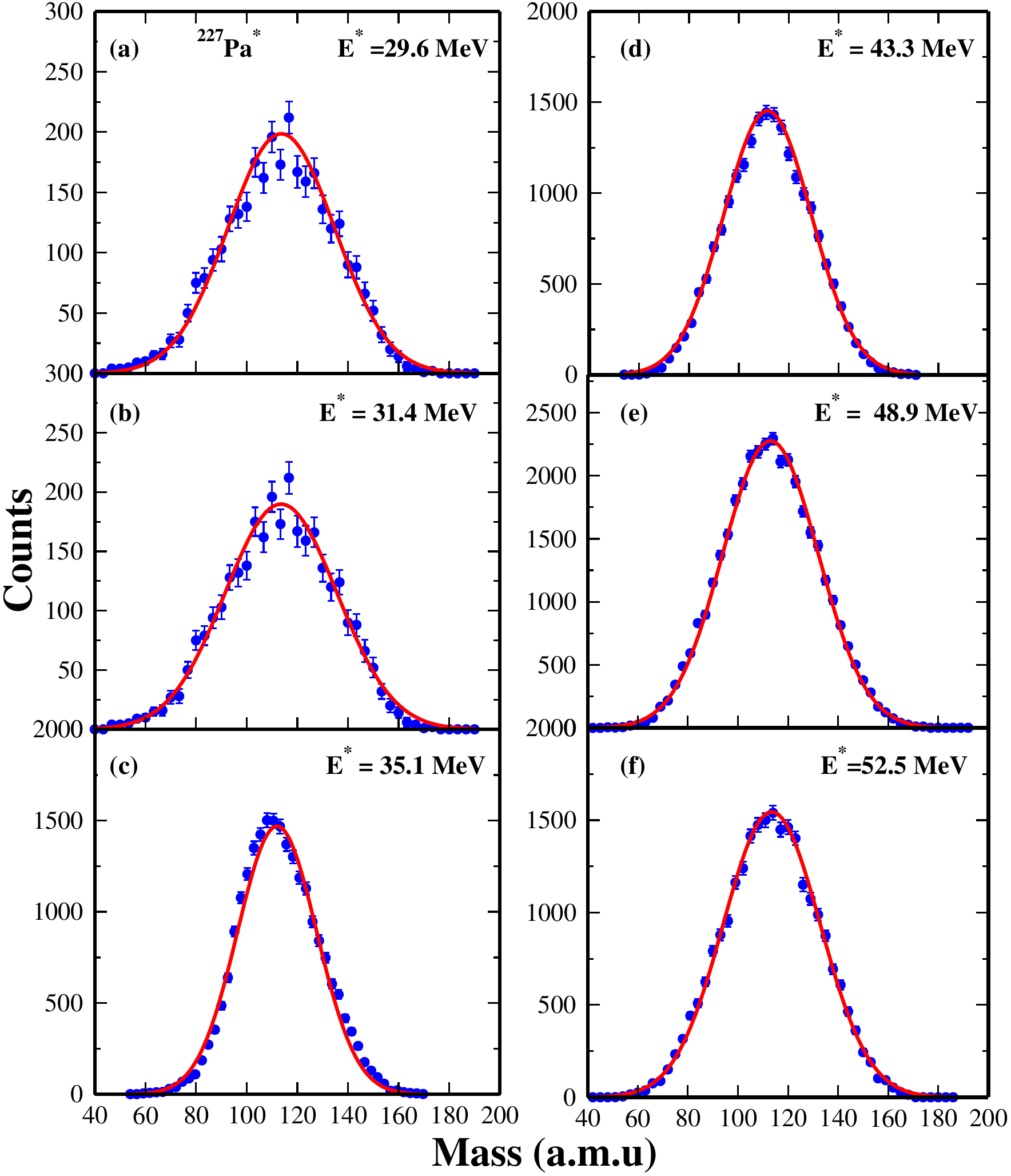}
\caption{\label{fig:Figure2} Measured mass distribution of fission fragments for the reaction $^{19}F$+$^{208}Pb$ $\longrightarrow$ $^{227}$Pa$^*$ at different excitation energies.}
\end{figure}

       No mass-angle correlation was observed in the reactions at energies below or above the fusion barrier indicating the absence of QF events. Fig. \ref{fig:Figure2} shows the FF mass distributions measured in the reaction $^{19}$F+$^{208}$Pb$\longrightarrow$$^{227}$Pa$^*$ at six excitation energies. Excitation energies are calculated after considering the energy loss of the beam in the target layer (half thickness) and carbon backing. The measured mass spectrum could well be described by a single symmetric Gaussian centered around mass $M\sim \frac{A_{CN}}{2}$. The shape of the mass distribution remains unchanged except at lower excitation energies where the mass distribution tends to deviate from symmetric Gaussian shape. Single Gaussian mass distribution feature shows the distinct signature of symmetric mass division of fission fragments. The standard deviation $\sigma_M$ of the mass distribution was obtained at each excitation energy after making best Gaussian fit to the data. As the excitation of compound nucleus increases, the $\sigma_M$ is found to increase. At lower excitation energies, the standard deviation from the fit shows more wider mass distribution than expected from the standard symmetric mass division of the FF. Similar analysis was done for the reaction  $^{19}$F+$^{206}$Pb $\longrightarrow$ $^{225}$Pa$^*$ measured up to same excitation energies and identical results were obtained. 
       
\begin{figure}[htb!]	
\centering
\includegraphics[width=1\linewidth]{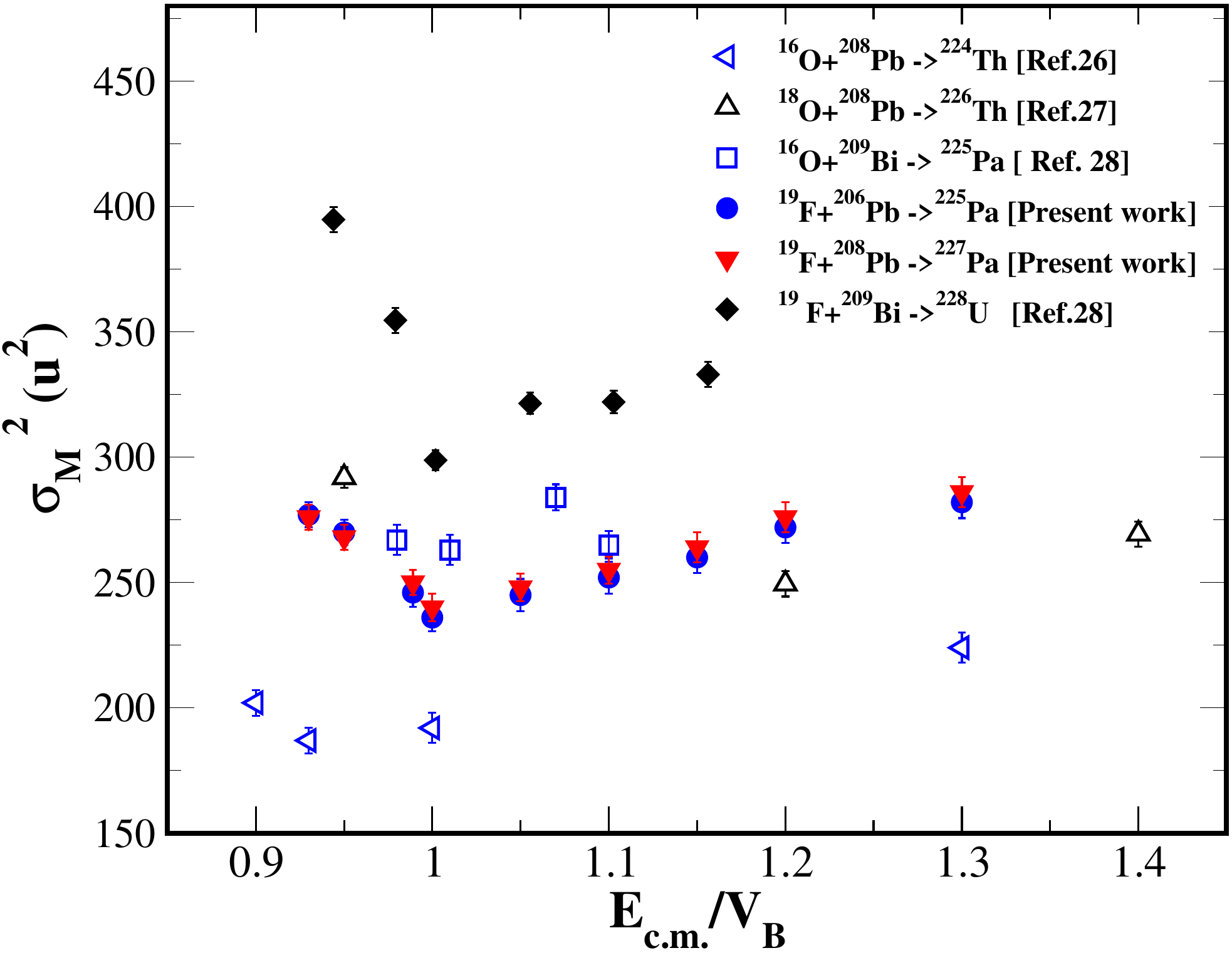}
\caption{\label{fig:Figure3}The mass variance  $\sigma_M^2$ for compound nucleus $^{225, 227}$Pa as a function of reduced bombarding energy E$_{c.m.}$/V$_{B}$. Data for nearby neutron-deficient nuclei  $^{224, 226}$Th, $^{228}$U \cite {M. Itkis95, Rusanov, Pant}  are also shown.}
\end{figure}

       The  mass variances  ($\sigma_M^2$)  as a function of E$_{c.m.}$/V$_{B}$ (where $V_{B}$ is the fusion barrier) for present systems are plotted in Fig. \ref{fig:Figure3}.  It is evident from the figure that, $\sigma_M^2$ decreases with decrease in beam energy till the fusion barrier and shows sudden increase below the barrier. The  $\sigma_M^2$ values for both systems  $^{19}$F+$^{206, 208}$Pb$\longrightarrow$$^{225, 227}$Pa$^*$ show similar trend. For comparison, we have also shown the variance distribution data deduced from the results of measurements on $^{225}$Pa and neighboring  $^{224, 226}$Th and $^{228}$U nuclei by other groups \cite {M. Itkis95, Rusanov, Pant}. In all reactions, variance of the mass distribution of fissioning nuclei demonstrate abrupt change below barrier energies. For the reaction $^{16}$O+$^{209}$Bi$\longrightarrow$$^{225}$Pa$^*$, similar trend can not be observed due to insufficient data points at lower energies \cite {Pant}. Though the relative values vary among different reactions, the nature of the variation is identical in all cases. This verifies that the sudden rise in $\sigma_M^2$ observed in our measurement appears to be a real fission property in this mass region.
       
\begin{figure}[htb!]	
\centering
\includegraphics[width=1\linewidth]{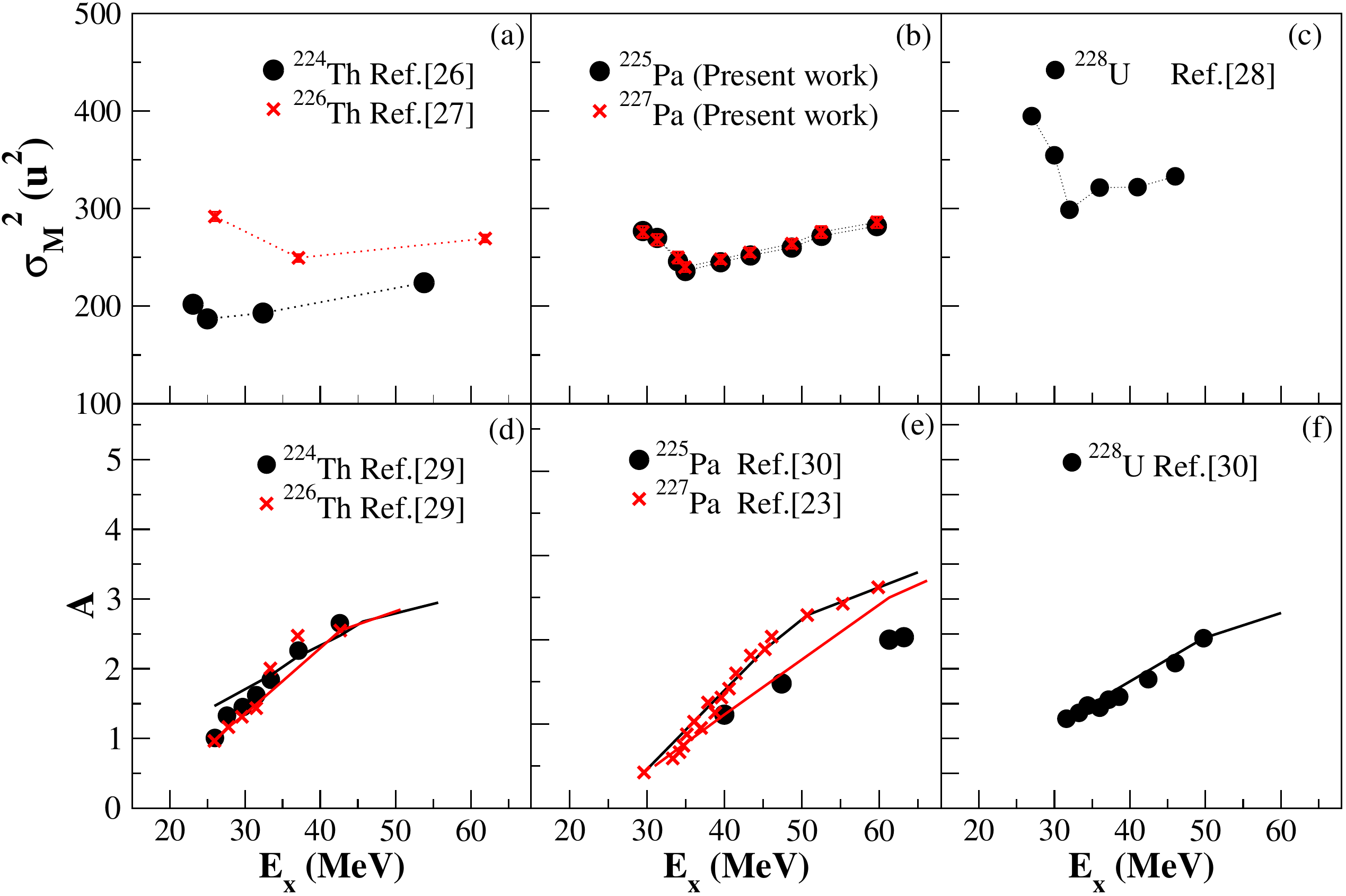}
\caption{\label{fig:Figure4} a-c) Variance of mass distribution and d-f) the angular anisotropy of FF from $^{224, 226}$Th, $^{225, 227}$Pa and $^{228}$U nuclei as function of CN  excitation energies  \cite {M. Itkis95, Rusanov, Pant, Vulgaris, Hinde1, Samant}. Solid lines in angular anisotropy (d-f) correspond to TSM calculation \cite {Vandenbosch}.}
\end{figure}

         Figure \ref{fig:Figure4} displays the variance of mass distribution and angular anisotropy (A)  of  FF from $^{225, 227}$Pa  nuclei  as well as nearby nuclei plotted as a function of the initial excitation energies. Data on angular anisotropies for nuclei Th, Pa and U were taken from \cite {Vulgaris, Hinde1, Samant}. All these reactions used spherical targets and were studied at energies similar to present measurement, i.e., below barrier to just above barrier energies. Table \ref{TABLE:1} shows the common reaction parameters for these systems. All of them have similar mass-asymmetry (mass-asymmetry lying below Businaro-Gallone point ($\alpha_{bg}$) \cite {Abe}), low Z$_{P}$Z$_{T}$ and  slightly varying fissility (${\chi}$) \cite{Blocki}. These reactions are expected to follow similar dynamics. As expected, all these systems show similar properties of fragment mass and angular distributions with respect to excitation energy. For all the reactions, as the CN excitation energy is decreased, the $\sigma_M^2$ values decreases monotonically, but shows sudden increase at lower excitation energies as shown in Fig.\ref{fig:Figure4}(a-c). From angular-distribution data, the angular anisotropy for all reactions (Fig.\ref{fig:Figure4}(d-e)) showed good agreement between experimental data and the predictions of transient statistical model (TSM) \cite {Vandenbosch} suggesting that the fission events followed from equilibrated compound nucleus  in all these systems.

\begin{table*}[htb]
\caption{Reaction parameters for various entrance channels populating  $^{224, 226}$Th, $^{225, 227}$Pa and $^{228}$U nuclei. Columns show the reaction, compound nucleus (CN),  Z$_{P}$Z$_{T}$, $\alpha_{bg}$, ${\alpha}( =\frac{A_{T} - A_{P}}{A_{T} + A_{P}}$), fissility (${\chi}$), and the neutron number N of CN.}
\label{TABLE:1}
\begin{tabular}{ccccccc}
 \hline
Reaction     &CN           &Z$_{P}$Z$_{T}$          &$\alpha_{bg}$  & ${\alpha}$  &${\chi}$   &N        \\
\hline
$^{16}$O + $^{208}$Pb         &$^{224}$Th               &656                         &0.871                &0.857        &0.763     &134     \\
$^{18}$O + $^{208}$Pb         &$^{226}$Th               &656                         &0.869                &0.840         &0.760     &136     \\
$^{19}$F + $^{206}$Pb         &$^{225}$Pa               &738                         &0.875                &0.831        &0.773     &134     \\
$^{16}$O + $^{209}$Bi         &$^{225}$Pa               &664                         &0.875                &0.858        &0.773     &134     \\
$^{19}$F + $^{208}$Pb         &$^{227}$Pa               &738                         &0.874                &0.832        &0.770     &136       \\
$^{19}$F + $^{209}$Bi         &$^{228}$U                &747                         &0.879                &0.833        &0.781     &136       \\[1ex]
\hline
\end{tabular}
\end{table*}

\section{DISCUSSIONS}

\subparagraph{}
      The two reactions, $^{19}$F on $^{206, 208}$Pb were performed at energy range of $0.93 \le \frac{E_{c.m}}{V_B}\le 1.29$. In neither reaction is QF expected at this energy range. If QF is present, it can influence the shape and width of mass distribution as reported in other works \cite {Toke85}. Though for reactions involving $^{16}$O and $^{19}$F projectiles on deformed targets, at lower energies broader mass distributions have been observed and attributed to presence of QF \cite {T.K.Ghosh}, no such evidence of QF has been reported in reactions with spherical target system such as $^{206, 208}$Pb at near-barrier energies. For the system $^{19}$F on $^{208}$Pb forming $^{227}$Pa, the existing experimental data on angular anisotropy measured around barrier energy showed that the reaction follows complete fusion-fission \cite {Hinde1}. Moreover, the theoretical work based on di-nuclear-system calculations performed by Nasirov \textit{et al.} also indicated that fusion-fission reaction dominates in the reaction $^{19}$F+$^{208}$Pb around barrier energies \cite{Nasirov}. Their theoretical results on the excitation function of the complete fusion and the angular anisotropy agreed well with the existing experimental data. The absence of any mass-angle correlation in our measurement and comparatively low Z$_{P}$Z$_{T}$  value of the system also suggests that contribution of QF is negligible in the present reactions. 

         The smooth variation of $\sigma_M^2$ with respect to excitation energy is the expected property of fission from complete fusion. Within the framework of liquid drop model, mass variance increases with excitation energy because of the temperature and angular momentum \cite {Ogn85}. Considering all of the fission events are due to complete fusion, the  sudden increase in $\sigma_M^2$ values at low excitation energies shows drastic changes in the fission property and may signify the presence of multi-mode fission in the system. The broader mass distribution at lower excitation of CN could be due to the superposition of mass distributions from two independent fission modes, one following the normal symmetric mode and the other following the asymmetric mode. Such admixture of fission modes had been clearly established in heavy-ion-induced fission of light actinide nuclei. Earlier measurements on mass and element yield distributions of fragments from fission of $^{225,227}$Pa nuclei formed in the reaction $^{16,18}$O + $^{209}$Bi reported presence of asymmetric fission components with light and heavy masses around mass numbers 90 and 137 respectively \cite{Nishinaka}. Relative yield of asymmetric component was about 10 \% of the total fission yield at 29 MeV of initial excitation energy. In the fission of $^{224,226}$Th nuclei produced through the reaction  $^{16,18}$O + $^{208}$Pb, multi-mode fission with more than one asymmetric fission components were observed below the barrier energies \cite {M. Itkis95, Rusanov}. Presence of four fission modes were realized in the fission of $^{226}$Th at 26 MeV of excitation energy \cite {Rusanov}. It was observed that a tin cluster with heavy fragment of mass A$\sim 140$ and light fragment close to spherical neutron shell with N $\sim$50 appear to be the stabilizing factor in the asymmetric fission mode \cite {Rusanov, Chubarian}. In a similar reaction $^{19}$F+$^{209}$Bi, mass distribution of $^{228}$U also showed rapid increase of mass variance below the barrier energies \cite {Pant}. In all these cases, the broader mass distributions at below-barrier energies were attributed to the contribution of asymmetric mass division manifested at lower excitation energies. The role of mass-asymmetric fission mode attributed to shell effects is clearly visible in all these reactions. Present work also shows similar results suggesting the contribution of asymmetric fission components in the fission of $^{225, 227}$Pa nuclei yielding to enhanced mass variance below-barrier energies. Fusion barrier for these systems corresponds to initial excitation energy of $\sim 35$ MeV. The observed broad mass widths below-barrier energies suggest that the shell effects persistent at low excitation energy($ < 35 $ MeV) could  influence the fission mode leading to increased contribution of asymmetric fission events.

       It should be noted that a significant fraction of the initial excitation energy is carried away by pre-fission particles so that the effective excitation energy at scission point is low \cite {M.G. Itkis91}.  In the studied systems, due to pre-fission neutrons, nuclei lighter than $^{225}$Pa may also be involved in fission process and for them the contribution of the asymmetric mode is smaller in low energy fission \cite {M. Itkis95, Rusanov}. Earlier experimental results suggested that  the transition from symmetric to asymmetric fission occur around A$_{CN}$=226 for excitation energy close to fission barrier\cite{Schmidt}. From our measurements, identical results from both reactions suggest that asymmetric components exist in both $^{225}$Pa (N=134) and $^{227}$Pa (N=136) and its influence on mass variance is clearly visible. This observation is in agreement with the results of Nishinaka \textit{et al.} where they observed asymmetric fission yields in both nuclei formed  in reaction $^{16, 18}$O + $^{209}$Bi$\longrightarrow$ $^{225,227}$Pa  \cite {Nishinaka}.
 
\begin{figure}[htb!]	
\centering
\includegraphics[width=0.6\linewidth]{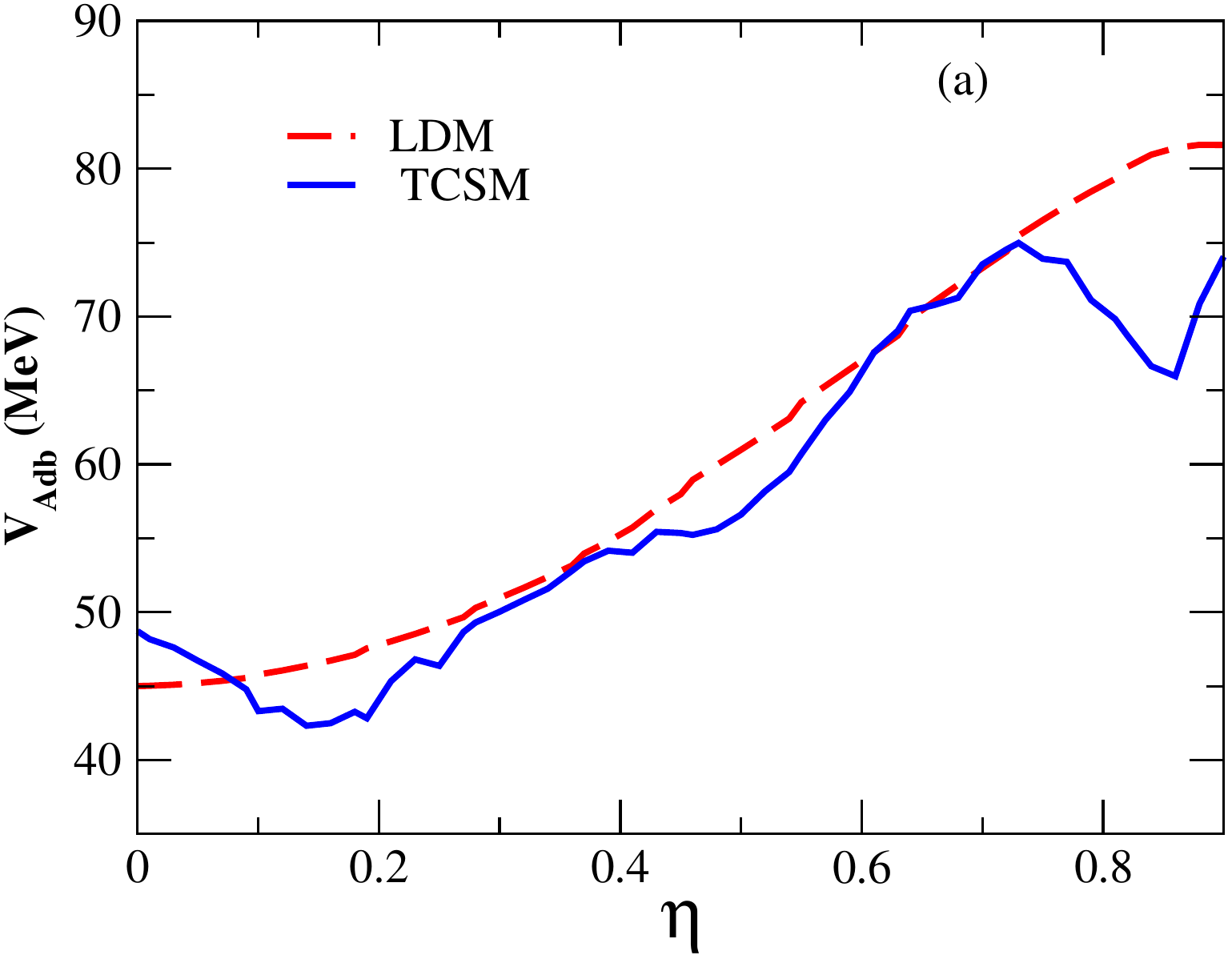}
\includegraphics[width=0.6\linewidth]{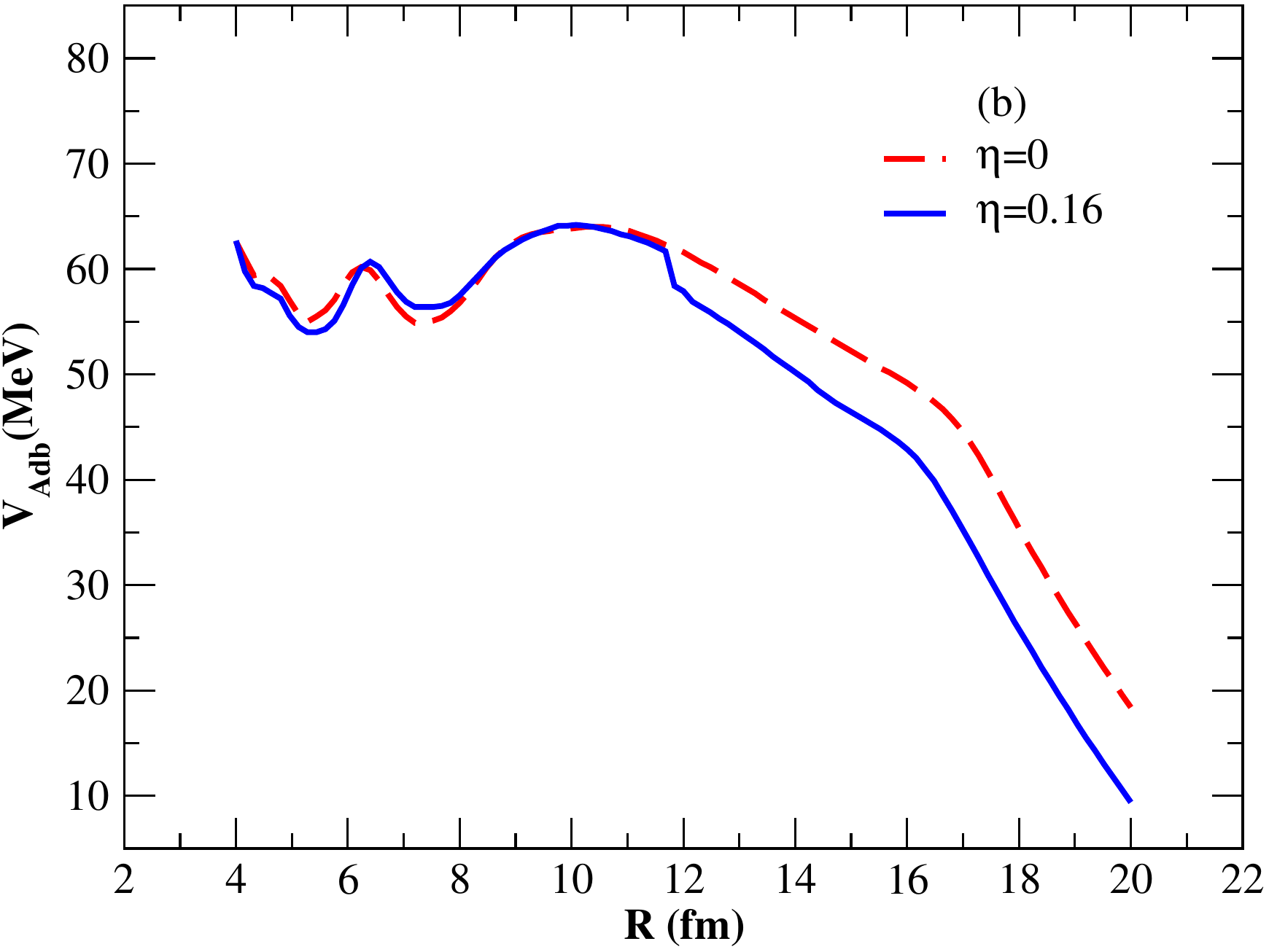}
\caption{\label{fig:Figure5} Calculated adiabatic potential energy for the fission of compound nucleus $^{227}$Pa. Plot a) $V_{Adb}$ near the scission point as a function of mass-asymmetry (${\eta}$) for LDM (dashed lines) and TCSM (continuous line),  and b) $V_{Adb}$ as a function of elongation R for $\eta = 0$ (dashed lines) and $\eta = 0.16$(continuous line). The potential energy calculation was performed with zero dynamical deformation ($\delta =0$) of nuclear fragments)(for explanation see the text)}
\end{figure}

      The presence of mass asymmetry in low energy fission has been verified  by multi-dimensional PES calculation of the fissioning system. In Fig. \ref{fig:Figure5}, we show the results of the multi-dimensional PES calculation performed for the fission of $^{227}$Pa formed in reaction  $^{19}$F + $^{208}$Pb using the standard macro-microscopic model based on Strutinsky shell correction method \cite {Strutinsky, V. M. Strutinsky,  M. Brack}. The macroscopic energy is calculated within the framework of finite-range liquid-drop model (LDM) and the shell correction is applied based on the well known two-center shell model(TCSM) as proposed in \cite {Zagrebaev, J. Maruhn}. The calculation of potential-energy landscape was performed in four-dimensional deformation space (R, ${\eta}$, $\delta$, $\epsilon$) where R is the distance between the center-of-mass points of two fragments, ${\eta}=\frac{m_2-m_1}{m_2+m_1}$ the mass-asymmetry parameter, $\delta$ the  unified dynamical deformation \cite {Greiner} and $\epsilon$ the neck parameter. The  recommended  value of $\epsilon$=0.35 was used such that the PES is minimal along the fission path \cite {Yamaji}. The adiabatic potential energies were calculated at the zero temperature (T=0) using the NRV code \cite{Zagrebaev}. 
 
      Figure \ref{fig:Figure5}a shows the calculated adiabatic potential energy ($V_{Adb}$) of deformed $^{227}$Pa nucleus near the scission point plotted  as a function of mass-asymmetry ${\eta}$ using LDM and TCSM respectively. Here one can see the potential valley located near asymmetry ${\eta}$ =0.16 (corresponds to mass division 132/95) which is energetically more favorable for fission. Clearly, the asymmetry of the fragments is influenced by the shell closure in the heavy fragments due to its proximity to doubly magic $^{132}$Sn (Z=50, N=82). In Fig. \ref{fig:Figure5}b we show the adiabatic potential energy as a function of elongation R for fission pathways corresponding to mass symmetry ($\eta=0$) and  mass asymmetry ($\eta=0.16$). Analysis of PES shows the symmetric and asymmetric mass split follows distinct fission paths having different energy thresholds and separated through saddle to scission. Multi-dimensional PES calculation by others also showed different valley structure of PES leading to existence of \emph{}multi-mode fission in light actinide nuclei at low excitation energies \cite {Moller01,Rusanov}.
      
\section{Conclusion}
  In conclusion, we have studied the multi-mode fission in light actinide nuclei via dependence of mass variance on excitation energy of the fissioning nucleus.  We observed that variance of the mass distribution increases at lower excitation energies where shell effects responsible for multi-mode fission become dominant.  As the excitation energy is increased, a transition from multi-mode fission to liquid-drop fission is observed. From the systematic analysis of $\sigma_M^2$ in light actinide nuclei, it is suggested that,  for fission induced by projectiles with mass $A_p < 20$ on spherical targets and  Z$_{P}$Z$_{T}$ ($ < 800$), the manifestation of asymmetric fission could broaden the mass distribution in these systems at low excitation energies. Contribution from QF is found to be negligibly small for these systems. In the present experiment, identical results for systems  $^{19}$F+ $^{206,208}$Pb $\longrightarrow$ $^{225,227}$Pa  suggests similar fission properties in both nuclei (N=134 and N=136) showing presence of asymmetric fission components influencing the mass variance.  The influence of shell effects on mass variance has been observed upto CN excitation energies of $\sim$ 35 MeV. Recent discovery of asymmetric mass division in fission of neutron deficient isotopes of $^{180, 190}$Hg showed the influence of shell effects persistent even upto initial excitation energy of  $\sim$70 MeV in $^{190}$Hg \cite {Andreyev, K. Nishio}. The dynamics and the parameters that govern the rate of shell damping as a function of reaction and excitation energy are still not fully understood \cite {Aritomo,J. Randrup}. More experiments in fission of neutron deficient transitional nuclei may offer important clues for better understanding of the complex fission process and its multi-modal nature.
  
\section{Acknowledgements}
We are thankful to accelerator group members at IUAC, New Delhi, especially R. Joshi and A. Sarkar, for their effort in providing stable pulsed beam throughout the experiment. One of the authors (RD) acknowledges the financial support from Council of Scientific Industrial Research, Govt. of India.

\bibliographystyle{model1a-num-names}
\bibliography{<your-bib-database>}

\end{document}